# 3D+1 PARAMETRIC BRIGHT VORTEX SOLITONS. OPTICAL MESONS.

Lubomir M. Kovachev

Institute of Electronics, Bulgarian Academy of Sciences

Tzarigradsko chaussee 72,1784 Sofia, Bulgaria

Abstract. Using the method of separation of variables, we developed a vector nonparaxial theory from the nonlinear wave equations in strong field approximation of a Kerr media. Investigating three waves on different frequencies, which satisfied the condition $2\omega_3 = \omega_1 + \omega_2$, we found that exact localized vortex parametric soliton solutions existed for the set of four nonlinear 3D+1 vector wave equations. The solutions are obtained for a fixed angle between the main frequencies and the signal waves $\alpha = 60^0$. This method is applicable for angular functions, which satisfy additional conditions. It is shown that exact parametric vortex solitary waves exist for solution with eigenrotation momentum L=1. The method is generalized for arbitrary number of signal waves appearing symmetrically in pairs in respect to $\omega_3$. The existence of such type of vortex solitons, in the case of degenerate four-wave mixing is discussed. As all of the waves are localized and have momentum L=1, they are called optical mesons.

1. Introduction.

Recently, [1] exact 3D+1 vortex soliton solutions of nonlinear wave equations for cubic media in an approximation for a strong field were obtained. It was possible to solve the vector equations, representing the vector of electrical field $\vec{E}$ as a sum of three orthogonal linear polarized components, namely $\vec{E} = E_x\vec{x} + E_y\vec{y} + E_z\vec{z}$. Investigating three localized wave packets on different frequencies, which satisfy the condition $2\omega_3 = \omega_1 + \omega_2$, additional nonlinear effects as Cross-Phase Modulation (CPM) and four-photon parametric mixing (FPM) appear. All these effects depend substantially on the polarization of the waves. In the simplest case, when all three waves are linear polarized in one direction, in one dimensional approximation (optical fiber or slab wave guides), the propagation of a slowly varying amplitude is described by solving of a Generalized System of Nonlinear Schredinger Equations (GSNSE). For this system parametric soliton solutions [2], symbiotic solitons [3] and periodic knoidal solutions [4] were obtained. Recently [5,6], some interesting numerical parametric soliton solutions for 1D+1 Helmholtz type equations were obtained. On the other hand, as it was pointed in [7,8] and experimentally demonstrated in [9] the CPM leads to acceleration between the waves and the existing of an interaction potential as a function of the distance between the weight centers of the localized waves. Practically, this interaction change the spectral weight of each wave, and makes them spectrally more closed. But as the deformation of the spectral shape due to CPM, is nonlinear and asymmetric to their mutual center of weight, the spectral weight of the wave is not equal to the spectral maximum in the interaction process. In this way, for close spectrally waves it is possible to obtain equal spectral weight for each wave (equal group velocity), and different spectral maximum of the waves. When the asymmetric deformation of the waves due to the CPM is quicker than their group delay, it is possible to obtain, so called mixed states of optical waves on different carrying frequencies. In regime of mixing the waves will be propagated without separation, due to the group velocity different, but the spectral maximum of the waves will be well separated. Exchange of energy, due to FPM also leads to an effect of self-confinement of the localized optical waves [10]. The analyze made in [8] shows that the soliton solutions for the GSNSE, in the case when the waves are full overlapped, are asymptotically free, and the waves move freely with their group velocity. A little displacement of the wave centers yields acceleration to their mutual center due to the CPM and the exchange of energy by FPM. The potential interaction, due to CPM and the exchange interaction due to FPM, appear also in 3D+1 case [11]. It is shown that for short distances between the waves, when they are well overlapped in the space, the behavior of the components has the characteristics of a strong interaction. The nonlinear theory of the light in 2D+1 dimension case, describes an additional nonlinear effect, self-focusing. The early theories [12-15] of propagation of an optical beam in nonlinear cubic media in 2D+1 case were based on investigation of scalar paraxial wave equations. These theories also require a linear polarization of the electrical field, which leads to a scalar theory. Recently, there has been a lot of interest [16-18] in the investigation of vortex solitary waves in nonlinear media. The study of the optical vortices was made numerically by using such type of scalar paraxial wave equations. As it was pointed out in [20,21], this theory was not valid for a very narrow beam, and a correct description of the beam behavior requires a vector analysis. Using an order of magnitude analysis method, it is not hard to see that all these results were obtained for wave packets in a weak nonlinear media. One quantitative criterion for weak nonlinearity is the dimensionless magnitude to be in order of:

$$\frac{\tilde{n}_2}{n_0^2}|E_0|^2 \approx \mu^2 \text{ ,where } \mu << 1.$$

The theory in this approximation was very well investigated in many books and papers, and it was expererimentally confirmed.

Presently, there are laser systems producing localized optical wave packets with magnitude in the order of:

$$\frac{\tilde{n}_2}{n_0^2}|E_0|^2 \approx 1 \qquad (1)$$

in suitable nonlinear media. For this media, condition (1) is the strong field approximation.

To make a correct analyze of propagation of optical waves in this approximation it is necessary to go back to the basic vector wave equation, or to the Maxwell equations for cubic nonlinear media.

2. The equation and order of magnitude analysis.

We started from the nonlinear wave equations derived from the Maxwell's equations of nonlinear cubic media:

$$-\Delta \vec{E} + \nabla(\nabla \cdot \vec{E}) + \frac{n_0^2}{c^2}\frac{\partial^2 \vec{E}}{\partial t^2} = -\frac{n_0^2}{c^2}\frac{\partial^2 \vec{P}_{nl}}{\partial t^2}. \qquad (2)$$

$$\vec{P}_{nl} = \frac{\tilde{n}_2}{n_0^2}(\vec{E} \cdot \vec{E}^*)\vec{E} \quad , \qquad (3)$$

where $n_0$ is the linear refractive index and $\tilde{n}_2$ is the Kerr coefficient.

As it was pointed out in [21], for an isotropic media, the frequency-degenerate nonlinear polarization $\vec{P}_{nl}$ of an arbitrary electric field $\vec{E}$ was always parallel to $\vec{E}$ and led to the fact that $\chi^{(3)}$ was a scalar. If we neglect the term $\nabla(\nabla \cdot \vec{E})$, we obtain the next vector wave equation from (2) and (3):

$$-\Delta \vec{E} + \frac{n_0^2}{c^2}\frac{\partial^2 \vec{E}}{\partial t^2} = -\frac{n_0^2}{c^2}\frac{\partial^2 \vec{P}_{nl}}{\partial t^2} \qquad (4)$$

It is well known that a spectral limited localized wave packet satisfies the conditions:

$$\Delta\omega \approx \mu.\omega_0; \Delta k \approx \mu.k_0, \mu << 1$$
$$\Delta x \sim \Delta y \sim \Delta z \sim r_0; r_0 \Delta k \sim 1; t_0 \Delta\omega \sim 1; \qquad (5)$$
$$n_0.\omega_0/c = k_0 \quad ; \quad \vec{E} = E_0.\vec{\tilde{E}} ,$$

where $k_0$ and $\omega_0$ are the carrying wave number and carrying frequencies of the wave packet, $\Delta\omega$ and $\Delta k$ are the frequencies and wave number spectral bandwidth, and $t_0, r_0$ are their temporal and spatial dimension. We performed the order of magnitude analysis of (4) for a localized wave packet bearing in mind relations (1) and (5). For dimensionless magnitudes we obtained:

$$\mu^2 k_0^2 \Delta\vec{\tilde{E}} \sim \mu^2 k_0^2 \frac{\partial^2 \vec{\tilde{E}}}{\partial t^2} \sim \mu^2 k_0^2 \frac{\partial^2 \vec{\tilde{P}}_{nl}}{\partial t^2}$$

These relations show that all three terms are of one order and for a strong field, which satisfies (1), we must solve the nonlinear wave equation (4).

3. Separate of the variables and vortex soliton solutions.

We have tried to solve (4) by using the method of separation of variables. The electric-field vector $\vec{E}(\vec{r},t)$ is supposed to be a sum of three linear polarized components:

$$\vec{E}(x,y,z,t) = (1/2i)\left(\vec{a}E_1(x,y,z)\exp(-i\omega_1 t) + \vec{b}E_2(x,y,z)\exp(-i\omega_2 t) + \vec{c}E_3(x,y,z)\exp(-i\omega_3 t)\right) - \text{c.c.} \quad (6)$$

which frequencies satisfy the relation: $2\omega_3 = \omega_1 + \omega_2$ and $\vec{a}, \vec{b}, \vec{c}$ are the linear polarization unique vectors for the corresponding wave. In a Cartesian coordinate system the wave equation (4), for three, linearly polarized components in arbitrary directions of type (6) is equal to a scalar system of nine nonlinear wave equations. The third-order susceptibility tensor has a representation on Cartesian basis, and we made an expansion of the polarization components of the field in such basis. For the polarization vectors (7) we choose a special Cartesian basic of type $\vec{a} = \vec{b} = \vec{x}$ and $\vec{x} \cdot \vec{c} = \cos\alpha; \vec{y} \cdot \vec{c} = \sin\alpha; \vec{z} \cdot \vec{c} = 0$. Practically, we make the expansion:

$$E_3\vec{c} = E_3\cos\alpha\vec{x} + E_3\sin\alpha\vec{y} = E_{3x}\vec{x} + E_{3y}\vec{y}$$

(7).

This expansion is well known from the experiments of Degenerate Four Wave Mixing (DFWM) [22].

In such basis (7), for three parametrically connected waves of type (6), the nonlinear wave equation (4) is reduced to four scalar equations of type of:

$$\left(\Delta E_1 + k_1^2 E_1 + n_2^1\left(|E_1|^2 + |E_2|^2 + |E_3|^2 + |E_2|^2 + \cos^2\alpha|E_3|^2\right)E_1 + \cos^2\alpha n_2^1 E_3^2 E_2^*\right)\vec{x} = 0$$

$$\left(\Delta E_2 + k_2^2 E_2 + n_2^2\left(|E_1|^2 + |E_2|^2 + |E_3|^2 + |E_1|^2 + \cos^2\alpha|E_3|^2\right)E_2 + \cos^2\alpha n_2^2 E_3^2 E_1^*\right)\vec{x} = 0 \quad (8),$$

$$\left(\Delta E_{3x} + k_3^2 E_{3x} + n_2^3\left(|E_1|^2 + |E_2|^2 + |E_3|^2 + |E_1|^2 + |E_2|^2\right)E_{3x} + 2E_1 E_2 E_{3x}^*\right)\vec{x} = 0$$

$$\left(\Delta E_{3y} + k_3^2 E_{3y} + n_2^3\left(|E_1|^2 + |E_2|^2 + |E_3|^2\right)E_{3y}\right)\vec{y} = 0$$

where $|\vec{k}_i|^2 = k_i^2 = \dfrac{n_0^2 \omega_i^2}{c^2}$ ; $n_2^i = k_i^2 \tilde{n}_2^i$ ;

We suggested that the waves be in the close spectral region. Then $n_0^i(\omega) = n_0 = \text{const}$ and $\Delta k = 2k_3 - k_2 - k_1 = 0$. When $\Delta k \neq 0$ with suitable change of dependant variables, it is possible to reduce the set (8) to a new set (8), but only with the new wave number $\tilde{k}_i^2 = k_i^2 \pm \Delta k_i^2/2$. The connection between the polarization and the coefficients in front of the cross-terms is clearly seen from (8). When all waves have one polarization (optical fiber and slab wave-guides), then $\cos\alpha = 1$ and we have the standard coefficient 2 in front of the cross-terms. When the waves are orthogonally polarized, then there are no parametric terms and the coefficients in front of the cross-terms are unity. As it was made in a previous paper [1] we would try to solve the system (8) by the method of separation of variables when additional conditions for the angular functions are satisfied, suggesting also, that $\alpha = 60^0$.

As (8) is a scalar system for the components, it can be written in spherical variables.

$$\frac{1}{r}\frac{\partial^2(rE_1)}{\partial r^2} + \frac{1}{r^2\sin\theta}\frac{\partial}{\partial\theta}\left(\sin\theta\frac{\partial E_1}{\partial\theta}\right) + \frac{1}{r^2\sin^2\theta}\frac{\partial^2 E_1}{\partial\varphi^2} + k_1^2 E_1 + n_2^1\left(|E_1|^2 + |E_2|^2 + |E_3|^2 + |E_2|^2 + \frac{1}{4}|E_3|^2\right)E_1 +$$
$$+ \frac{1}{4}n_2^1 E_3^2 E_2^* = 0$$

$$\frac{1}{r}\frac{\partial^2(rE_2)}{\partial r^2} + \frac{1}{r^2\sin\theta}\frac{\partial}{\partial\theta}\left(\sin\theta\frac{\partial E_2}{\partial\theta}\right) + \frac{1}{r^2\sin^2\theta}\frac{\partial^2 E_2}{\partial\varphi^2} + k_2^2 E_2 + n_2^1\left(|E_2|^2 + |E_1|^2 + |E_3|^2 + |E_1|^2 + \frac{1}{4}|E_3|^2\right)E_2 +$$
$$+ \frac{1}{4}n_2^1 E_3^2 E_1^* = 0 \qquad (9)$$

$$\frac{1}{r}\frac{\partial^2(rE_{3x})}{\partial r^2} + \frac{1}{r^2\sin\theta}\frac{\partial}{\partial\theta}\left(\sin\theta\frac{\partial E_{3x}}{\partial\theta}\right) + \frac{1}{r^2\sin^2\theta}\frac{\partial^2 E_{3x}}{\partial\varphi^2} + k_3^2 E_{3x} + n_2^3\left(|E_1|^2 + |E_2|^2 + |E_3|^2 + |E_1|^2 + |E_2|^2\right)E_{3x} +$$
$$+ 2n_2^3 E_1 E_2 E_{3x}^* = 0$$

$$\frac{1}{r}\frac{\partial^2(rE_{3y})}{\partial r^2} + \frac{1}{r^2\sin\theta}\frac{\partial}{\partial\theta}\left(\sin\theta\frac{\partial E_{3y}}{\partial\theta}\right) + \frac{1}{r^2\sin^2\theta}\frac{\partial^2 E_{3y}}{\partial\varphi^2} + k_3^2 E_y + n_2^3\left(|E_1|^2 + |E_2|^2 + |E_3|^2\right)E_{3y} = 0$$

This is a set of four scalar nonlinear Helmholtz equations, and bearing in mind, that the $\bar{y}$ component for the $E_3$ waves has not parametric, we represented the components of the field as a product of radial and an angular part:

$$E_i = R_i(r)Y_i(\theta,\varphi), \quad i = 1,2,3 \qquad (14)$$

with the additional condition of the angular parts:

$$|Y_1|^2 + |Y_2|^2 + |Y_3|^2 + |Y_2|^2 + \frac{1}{4}|Y_3|^2 + \frac{1}{4}Y_3^2 = \text{conct}$$
$$|Y_2|^2 + |Y_1|^2 + |Y_3|^2 + |Y_1|^2 + \frac{1}{4}|Y_3|^2 + \frac{1}{4}Y_3^2 = \text{const} \qquad (10)$$
$$|Y_1|^2 + |Y_2|^2 + |Y_3|^2 + |Y_1|^2 + |Y_2|^2 + 2Y_1 Y_2 = \text{const}$$
$$|Y_1|^2 + |Y_2|^2 + |Y_3|^2 = \text{const.}$$

The nonlinear coefficients and wave numbers satisfy the correlations:

$$n_2^1 k_1^2 = n_2^2 k_2^2 = n_2^3 k_3^2 = n_2$$

Multiplying each equation of (9) with corresponding $\frac{r^2}{R_i \cdot Y_i}$, $i = 1,2,3$, and bearing in mind condition (10) we obtained:

$$\frac{r^2 \nabla_r^2 R_i}{R_i} + r^2\left(k_i^2 + c.n_2|R|_i^2\right) = -\frac{\nabla_{\theta,\varphi}^2 Y_i}{Y_i} = l(l+1) \qquad i = 1,2,3 \qquad (11)$$

where l is a number and

$$\nabla_r^2 = \frac{1}{r^2}\frac{\partial}{\partial r}\left(r^2 \frac{\partial}{\partial r}\right) \qquad (12)$$

$$\nabla_{\theta,\varphi}^2 = \frac{1}{\sin\theta}\frac{\partial}{\partial\theta}\left(\sin\theta\frac{\partial}{\partial\theta}\right) + \frac{1}{\sin^2\theta}\frac{\partial^2}{\partial\varphi^2}. \qquad (13)$$

are correspondingly the radial and the angular operators. In this way, the following equations for the radial and the angular parts of the wave functions are obtained:

$$\nabla_r^2 R_i + k_i^2 R_i + cn_2|R|_i^2 R_i - \frac{l(l+1)}{r^2}R_i = 0 \qquad (14)$$

$$\nabla_{\theta,\varphi}^2 Y_i + l(l+1)Y_i = 0 \qquad (15)$$

The substitution (11) with condition (10) allow us to separate the nonlinear part in the radial component of the fields and for the angular parts we have a usual linear eigenvalue problem. Equations (15) are well known equations and each of them has exact solutions of kind:

$$Y_i = Y_l^m(\theta,\varphi) = \Theta_l^m \Phi_m = \sqrt{\frac{2l+1}{4\pi}\frac{(l-m)!}{(l+m)!}} P_l^m(\cos\theta)e^{im\varphi} \qquad i = x,y,z. \qquad (16)$$

where $P_l^m$ are the Legendre's functions for discrete series of numbers:

$l = 0,1,2,...; m = 0,\pm 1,\pm 2,...$ and $|m| \leq l$.

Going back to set (9), we can see that it is only possible to separate the variables for the spherical functions, which satisfy condition (10). It is not hard to see that condition (10) is satisfied only by eigenfunctions with $l=1, m = 0, \pm 1$, namely:

$$Y_1 = Y_1^{-1} = -\sqrt{\frac{3}{8\pi}} e^{-i\varphi} \sin\theta;$$
$$Y_2 = Y_1^1 = \sqrt{\frac{3}{8\pi}} e^{i\varphi} \sin\theta; \qquad (17)$$
$$Y_3 = Y_1^0 = \sqrt{\frac{3}{4\pi}} \cos\theta;$$

By choosing each field components, one of these angular components (10) is satisfied, and we see that the eigenfunctions (16) are solutions for the angular part of the set of equations (9). The radial part of equation (14) has "de Broglie solitons" [23,24] solutions.

$$R_i = \frac{1}{2i} \frac{\sqrt{2}}{\sqrt{c.n_2^i}} \left( \frac{e^{ik_i r}}{r} - c.c. \right)$$

In this way, the existence of the localized parametric vortex soliton solutions for the vector wave equation with eigenrotation momentum l=1 has been found, and for a fixed choice of the polarization (7) it is:

$$\vec{E} = -\frac{\sqrt{2}}{2i} \sqrt{\frac{3}{8\pi.c.n_2^1}} \left( \frac{e^{i(k_1 r - \omega_1 t)}}{r} e^{-i\varphi} - c.c. \right) \sin\theta \vec{a} + \frac{\sqrt{2}}{2i} \sqrt{\frac{3}{8\pi.c.n_2^2}} \left( \frac{e^{i(k_2 r - \omega_2 t)}}{r} e^{i\varphi} - c.c. \right) \sin\theta \vec{b} +$$
$$+ \frac{\sqrt{2}}{2i} \sqrt{\frac{3}{4\pi.c.n_2^3}} \left( \frac{e^{i(k_3 r - \omega_3 t)}}{r} - c.c. \right) \cos\theta \vec{c} \qquad (19)$$

The module of this vector is:

$$|\vec{E}| = \sqrt{|\vec{E}|^2} = \sqrt{\frac{1}{2\tilde{n}_2} \frac{\sin^2 k_1 r}{(k_1 r)^2} + \frac{1}{2\tilde{n}_2} \frac{\sin^2 k_2 r}{(k_2 r)^2} + \frac{1}{\tilde{n}_2} \frac{\sin^2 k_3 r}{(k_3 r)^2}} \qquad (18)$$

From (18) it is seen that the soliton solution is localized and has finite value when $r \to 0$. The localized wave solutions are polarized in x-y plane. The stability of such type of solutions will be investigated in a next paper. But there is one additional possibility to compensate self-focussing of the field by one rotation with orbital momentum l=1. Thus we would obtain stable vortex optical solitons. A full analysis of the stability of the vector soliton solutions of a set of equations (9) must be made using the Hamiltonian formalism and getting the fact of appearing of the additional invariant connected with the momentum. It is not hard to see that such type of solutions exist also in the case of DFWM when:

$$\omega_1 = \omega_2 = \omega_3; k_1 = \pm k_2; \vec{k}_{1,2} \cdot \vec{k}_3 = \cos\alpha$$

4. Optical mesons.

The experimental fact that signal waves appear while the increasing of the intensity of the electrical field symmetrically on the both side of the main frequency $\omega_0$, due to FPM is well known. They appear in pairs, and are connected with the modulation instability of the basic waves. For example, after the 3 wave four- photon interaction, 5 wave pour photon interactions appear satisfying the conditions:

$$2\omega_3 = \omega_1 + \omega_2; 2\omega_3 = \omega_4 + \omega_5;$$
$$\omega_1 + \omega_2 = \omega_4 + \omega_5;$$

In every next step, while increasing of the wave intensity, two additional signal waves will appear. For an arbitrary number of waves, the next type of conditions for the waves will be satisfied:

$$2\omega_0 = \omega_{2i-1} + \omega_{2i}; \omega_{2i-1} + \omega_{2i+1} = \omega_{2i} + \omega_{2i+2}; ... i=1,2,...n.$$

For components, linearly polarized in different directions, this leads to a system of N coupled nonlinear parametric Helmholtz equations. It is well known fact from the theory of spherical functions, that:

$$\sum_{m=0}^{2l+1} \left| Y_l^m \right|^2 = \text{const} \qquad (21)$$

By using both optical components with suitable polarization and condition (21) for the spherical parts, it is possible to solve a system of N coupled parametric Helmholtz equations. The angular parts of these localized solutions is obtained by solving a linear eigenvalue problem, for eigenvalue l=1,2..n. That is why, we call these localized optical solitons optical mesons.

5. Conclusion.

An order of magnitude analysis is made for the nonlinear wave equations in strong field approximation. Using the method of separation of variables, it is shown that exact localized parametric vortex solitary solutions exist with eigenrotation momentum l=1. In contrast to the standard linear scalar theories, where the stationary solutions can be obtained for an infinite discrete series of numbers $k^2, l=0,1,2,...m=0,\pm 1,\pm 2,.....$, the nonlinear parametric equations admit finite number of soliton solutions, because of the requirement that the angular parts of equations should satisfy additional relations.